\documentclass[a4paper,11pt]{article}
\pdfoutput=1 

\usepackage{jcappub} 

\title{\boldmath Neutron stars in $f(R,T)$ gravity with conserved energy-momentum tensor: Hydrostatic equilibrium and asteroseismology}

\author[a]{Juan M. Z. Pretel, }
\author[a]{Sergio E. Jor\'as, }
\author[a,b]{Ribamar R. R. Reis}
\author[c,d]{and Jos\'e D. V. Arba\~nil}

\affiliation[a]{Instituto de F\'\i sica, Universidade Federal do Rio de Janeiro, \\ CEP 21941-972 Rio de Janeiro, RJ, Brazil}
\affiliation[b]{Observat\'orio do Valongo, Universidade Federal do Rio de Janeiro, \\ CEP 20080-090 Rio de Janeiro, RJ, Brazil}

\affiliation[c]{Departamento de Ciencias, Universidad Privada del Norte, \\ Avenida el Sol 461 San Juan de Lurigancho, 15434 Lima, Peru}
\affiliation[d]{Facultad de Ciencias F\'isicas, Universidad Nacional Mayor de San Marcos, \\ Avenida Venezuela s/n Cercado de Lima, 15081 Lima, Peru}

\emailAdd{juanzarate@if.ufrj.br}
\emailAdd{joras@if.ufrj.br}
\emailAdd{ribamar@if.ufrj.br}
\emailAdd{jose.arbanil@upn.pe}

\abstract{ We investigate the equilibrium and radial stability of spherically symmetric relativistic stars, considering a polytropic equation of state (EoS), within the framework of $f(R,T)$ gravity with a conservative energy-momentum tensor. Both modified stellar structure equations and Chandrasekhar's pulsation equations are derived for the $f(R,T)= R+ h(T)$ gravity model, where the function $h(T)$ assumes a specific form in order to safeguard the conservation equation for the energy-momentum tensor. The neutron star properties, such as radius, mass, binding energy and oscillation spectrum are studied in detail. Our results show that a cusp --- which signals the appearance of instability --- is formed when the binding energy is plotted as a function of the compact star proper mass. We find that the squared frequency of the fundamental vibration mode passes through zero at the central-density value corresponding to such a cusp where the binding energy is a minimum. }

\begin{document}
\maketitle
\flushbottom

\section{Introduction}\label{Sect1}

Extensions of General Relativity (GR) have been pursued \cite{PhysRevD.16.953} before the advent of the inflationary model and before any actual need for dark energy, since there was no data on the current accelerated expansion. In order to explain either one (or both of them), one of the most widely investigated modifications is the generalization of the original Einstein-Hilbert Lagrangian to a non-linear function of the Ricci scalar: $f(R)$ \cite{SotiriouFaraoni, Felice}.

A further generalization of $f(R)$ extended theories of gravity was developed by Harko and collaborators \cite{Harko2011} in order to introduce an unusual coupling between geometry and matter, namely $f(R,T)$ gravity. The inclusion of the energy-momentum trace $T$ may be justified by quantum effects or by, in its most general formulation, the lack of energy-momentum conservation \cite{Harko2011, LobatoEPJP2019}. 

The cosmological and astrophysical consequences of some $f(R,T)$ gravity models have been explored over the last few years. As a matter of fact, the cosmological solutions of $f(R,T)$ gravity for a perfect fluid in a spatially Friedmann-Lema{\^i}tre-Robertson-Walker metric in phase space were investigated in Ref. \cite{Shabani2013}, and the evolution of scalar cosmological perturbations was studied in Ref. \cite{Alvarenga2013}. The post-Newtonian gamma parameter in such theories of gravity was obtained in Ref. \cite{Shabani2014} and it has been shown that the mass corresponding to the interaction term leads to a flat rotation curve in the halo of galaxies \cite{Zaregonbadi2016}. Other more recent studies on cosmological applications of $f(R,T)$ gravity (including in the inflationary scenario) can be found in Refs. \cite{Xu2016, Moraes2017, Shabani2018, Debnath2019, Bhattacharjee2020, Bhattacharjee2020EPJP, Gamonal2021}. To the best of our knowledge, none of them have presented so strong constraints on the amplitude of the modification as the present work.

Astrophysical phenomena such as dissipative gravitational collapse have also recently been investigated within the context of such theories. Indeed, the effect of charge on non-adiabatic gravitational collapse dynamics in $f(R,T)$ gravity using the Misner-Sharp approach was analyzed in Ref. \cite{Ahmed2020}. Moreover, the dynamical equations for the spherical collapse in the form of heat flux, free-streaming radiation and shearing viscosity were presented in Ref. \cite{Sarbari2021}. In both works a causal heat transport equation from the M{\"u}ller-Israel-Stewart theory for dissipative fluids was used.

Due to its simplicity, the most studied model of $f(R,T)$ gravity at astrophysical level has been the particular functional form $f(R,T)= R+ 2\beta T$ with $\beta$ being a constant. Static compact stars solutions using isotropic perfect fluids were found in Refs. \cite{Moraes2016, Das2016, Deb2018, Lobato2020, Pretel2021}, while other authors have studied the effect of anisotropy on the compact star structure in the framework of this modified gravity model \cite{Deb2019, Maurya2019, Deb_2019JCAP, Biswas2020, Maurya2020, Rej2021, BISWAS2021168429}. Nevertheless, there is a peculiar feature in all these works because the energy-momentum tensor is a non-conserved quantity. This led some authors to investigate equilibrium configurations in $f(R,T)$ gravity under the requirement that the covariant derivative of the energy-momentum tensor is zero, see for instance Refs. \cite{Santos2019, Carvalho2020}.

On one hand, the fact that the four-divergence of the energy–momentum tensor is non-zero opens the possibility of a gravitationally induced particle production as shown by some authors \cite{Harko2014, Harko2015}. For further discussions on the non-conservation of energy-momentum tensor, see  Refs. \cite{Bertolami2008, Harko2008, Bisabr2012, Minazzoli2013}. However, the authors in Ref. \cite{Santos2019} argue that the creation or destruction of matter particles would not occur in a static analysis as in equilibrium compact stars. In particular, for the $f(R,T)= R+ h(T)$ gravity model, they shown that $h(T)$ must assume a specific form in order that $\nabla_\mu T^{\mu\nu}= 0$. Taking this study as a reference, our proposal will be to analyze the stellar stability.

The principal aim of this paper is to show that it is possible to construct stable compact stars in the framework of $f(R,T)$ gravity with conserved energy-momentum tensor. To do so, our first task is to generate a sequence of equilibrium configurations by demanding that the four-divergence of the energy-momentum tensor is zero and by solving the modified TOV equations. Our second step is to perturb all equations to first order in the metric and fluid variables following a procedure analogous to that carried out by Chandrasekhar in GR \cite{Chandrasekhar}. Once the oscillation equations are obtained, we then proceed to calculate the frequencies of the normal radial modes in order to analyze the stellar stability.

The present work is organized as follows: In Sec. \ref{Sect2} we briefly review $f(R,T)$ gravity, we explicitly express the field equations for the $f(R,T)= R+ h(T)$ model given a stellar spherically symmetric system, we introduce the different definitions of mass and we derive the modified TOV equations. In Sec. \ref{Sect3} we study the radial stability by considering small perturbations from the equilibrium state, and we derive the pulsation equations. Section \ref{Sect4} presents two polytropic EoSs widely used in the literature to describe neutron stars. In Sec. \ref{Sect5} we discuss our numerical results for the equilibrium configurations as well as analyze their radial stability. Finally, our conclusions are presented in Sec. \ref{Sect6}. It is worth mentioning that we will use a geometric unit system and the sign convention $(-,+,+,+)$. However, we chose to show our results in physical units to ease comparisons with others in the literature.


\section{Stellar structure equations}\label{Sect2}

In order to study compact stars, we present here the main aspects of stellar structure equations within the framework of $f(R,T)$ gravity. In $f(R,T)$ theories of gravity, the Einstein-Hilbert action of GR is modified by using a general function of $R$ and $T$, the Ricci scalar and the trace of the energy-momentum tensor, respectively. The action is given by 
\begin{equation}\label{1}
    S = \int d^4x\sqrt{-g}\left[ \frac{1}{16\pi}f(R,T) + \mathcal{L}_m \right] ,
\end{equation}
where $\mathcal{L}_m$ is the Lagrangian density for the matter distribution and $g$ is the determinant of the metric $g_{\mu\nu}$.

\subsection{Field equations}

From the variation of the action (\ref{1}) with respect to the metric tensor, one gets the field equations in the metric formalism \cite{Harko2011}
\begin{equation}\label{2}
f_RR_{\mu\nu} - \dfrac{1}{2}g_{\mu\nu}f + [g_{\mu\nu}\square - \nabla_\mu   \nabla_\nu] f_R  = 8\pi T_{\mu\nu} - (T_{\mu\nu} + \Theta_{\mu\nu})f_T ,
\end{equation}
where $f_R \equiv \partial f(R,T)/\partial R$, $f_T \equiv \partial f(R,T)/\partial T$, $\square \equiv \nabla_\mu\nabla^\mu$ is the d'Alembert operator with $\nabla_\mu$ representing the covariant derivative, and the tensor $\Theta_{\mu\nu}$ is defined as 
\begin{equation}\label{3}
\Theta_{\mu\nu} \equiv g^{\alpha\beta}\frac{\delta T_{\alpha\beta}}{\delta g^{\mu\nu}}  = -2T_{\mu\nu} + g_{\mu\nu}\mathcal{L}_m - 2g^{\alpha\beta} \frac{\partial^2\mathcal{L}_m}{\partial g^{\mu\nu} \partial g^{\alpha\beta}} .
\end{equation}

By taking the covariant derivative of Eq. (\ref{2}), one can obtain the following modified equation for the four-divergence of the energy-momentum tensor \cite{Barrientos2014}
\begin{equation}\label{4}
\nabla^\mu T_{\mu\nu} = \frac{f_T}{8\pi - f_T}\left[ (T_{\mu\nu} + \Theta_{\mu\nu})\nabla^\mu \ln f_T  + \nabla^\mu\Theta_{\mu\nu} - \frac{1}{2}g_{\mu\nu}\nabla^\mu T \right] ,
\end{equation}
and the trace of the field equations (\ref{2}) leads to
\begin{equation}\label{5}
3\,\square f_R + Rf_R - 2f = 8\pi T - (T+\Theta)f_T .
\end{equation}

Since there is no unique definition of the matter Lagrangian density, in the present study we assume that $\mathcal{L}_m = -\rho$; where $\rho$ stands for the energy density. Then, Eq. (\ref{3}) implies that 
\begin{equation}\label{6}
\Theta_{\mu\nu} = -2T_{\mu\nu}- \rho g_{\mu\nu} ,
\end{equation}
so that the field equation (\ref{2}) becomes 
\begin{equation}\label{7}
  f_RR_{\mu\nu} - \dfrac{1}{2}g_{\mu\nu}f + [g_{\mu\nu}\square - \nabla_\mu\nabla_\nu] f_R  = 8\pi T_{\mu\nu} + (T_{\mu\nu} +\rho g_{\mu\nu})f_T .
\end{equation}

In addition, following Ref. \cite{Santos2019}, we consider a particular case of $f(R,T)$ theories in which the function $f$ is given by $f(R,T)= R+ h(T)$, where $h(T)$ is an arbitrary function of the trace of the matter energy-momentum tensor. Thus, $f_R= 1$, $f_T = dh/dT \equiv h_T$, and Eqs. (\ref{7}), (\ref{4}) and (\ref{5}) take the following form, respectively, 
\begin{subequations}
\begin{align}
  G_{\mu\nu} &= 8\pi T_{\mu\nu} + \frac{1}{2}hg_{\mu\nu} + (T_{\mu\nu} + \rho g_{\mu\nu})h_T , \label{8a}   \\
  \nabla^\mu T_{\mu\nu} &= -\frac{h_T}{8\pi + h_T}\left[ (T_{\mu\nu}+ \rho g_{\mu\nu})\nabla^\mu\ln h_T  + \nabla_\nu\left( \rho+ \frac{1}{2}T \right) \right],  \label{8b}   \\
  R &= -8\pi T -2h -(T+ 4\rho)h_T ,  \label{8c}  
\end{align}
\end{subequations}
where $G_{\mu\nu}$ is the Einstein tensor. If the energy-moment tensor is a conserved quantity $\nabla^\mu T_{\mu\nu} = 0$, then we must demand that
\begin{equation}\label{9}
(T_{\mu\nu}+ \rho g_{\mu\nu})\nabla^\mu\ln h_T + \nabla_\nu\left( \rho+ \frac{1}{2}T \right) = 0 .
\end{equation}

From now on we will only consider that spacetime geometry is described by the spherically symmetric metric
\begin{equation}\label{10}
ds^2 = -e^{2\psi}dt^2 + e^{2\lambda}dr^2 + r^2(d\theta^2 + \sin^2\theta d\phi^2) ,
\end{equation}
where $x^\mu = (t, r, \theta, \phi)$ are the Schwarzschild  coordinates, and the metric potentials $\psi$ and $\lambda$ depend on both coordinates $t$ and $r$.

To describe the stellar matter source we adopt the energy-momentum tensor of an isotropic perfect fluid, namely
\begin{equation}\label{11}
T_{\mu\nu} = ( \rho + p )u_\mu u_\nu + pg_{\mu\nu} ,
\end{equation}
where $p$ represents the pressure of the fluid, and $u^\mu$ stands for the four-velocity of an observer comoving with the fluid which is given by
\begin{equation}\label{12}
u^\mu = \dfrac{dx^\mu}{d\tau} = u^0\left( 1, \dfrac{dx^i}{dt} \right) ,
\end{equation}
with $\tau$ being the proper time. Since we are considering a spherically symmetric system with motions, if any, only in the radial directions, then $u^\mu = (u^0, u^1, 0, 0)$ and $T_2^{\ 2} = T_3^{\ 3} = p$.

Consequently, for the line element (\ref{10}) and energy-momentum tensor (\ref{11}), we obtain
\begin{equation}\label{13}
\nabla_\nu T_1^{\ \nu} = \partial_t T_1^{\ 0} + \partial_r T_1^{\ 1} + T_1^{\ 0}(\dot{\psi} + \dot{\lambda}) + \left[T_1^{\ 1} - T_0^{\ 0}\right]\psi'  + \dfrac{2}{r}\left[T_1^{\ 1} - p\right] = 0 , 
\end{equation}
where overdots and primes denote partial differentiation with respect to $t$ and $r$, respectively. The non-zero components of the field equations (\ref{8a}) are given by 
\begin{align}
 &\dfrac{1}{r^2}\partial_r(re^{-2\lambda}) - \dfrac{1}{r^2} = 8\pi T_0^{\ 0} +\frac{1}{2}h + (T_0^{\ 0} + \rho)h_T ,   \label{14}   \\  
 &e^{-2\lambda}\left( \dfrac{2}{r}\psi' + \dfrac{1}{r^2} \right) - \dfrac{1}{r^2} = 8\pi T_1^{\ 1} +\frac{1}{2}h + (T_1^{\ 1} + \rho)h_T ,   \label{15}    \\
 &e^{-2\lambda}\left[ \psi'' + \psi'^2 - \psi'\lambda' + \dfrac{1}{r}(\psi' - \lambda') \right]  + e^{-2\psi}\left[ \dot{\lambda}\dot{\psi} - \ddot{\lambda} - \dot{\lambda}^2 \right] = 8\pi p +\frac{1}{2}h+ (\rho+ p)h_T ,  \label{16}    \\
 &\dfrac{2}{r}e^{-2\lambda}\dot{\lambda} = (8\pi + h_T)T_0^{\ 1} ,  \label{17}
\end{align}
and, in view of Eqs. (\ref{14}) and (\ref{15}), we can obtain a useful relation which will be used later: 
\begin{equation}\label{18}
\frac{2}{r}e^{-2\lambda}(\psi'+\lambda') = (8\pi+ h_T)(T_1^{\ 1} - T_0^{\ 0}) .
\end{equation}

\subsection{Modified TOV equations}

The metric and thermodynamic quantities do not depend on time when the star remains in a state of hydrostatic equilibrium so that $T_0^{\ 0} = -\rho_0$ and $T_1^{\ 1} = T_2^{\ 2} = T_3^{\ 3} = p_0$. Here the values corresponding to such equilibrium are denoted by a lower index $0$. Thus, the field equations (\ref{14}) and (\ref{15}) become
\begin{eqnarray}
  \frac{d}{dr}(re^{-2\lambda_0}) &=& 1- 8\pi \rho_0r^2 + \frac{r^2}{2}h_0 , \label{19}\\
  \frac{2}{r}e^{-2\lambda_0}\psi'_0 &=&\frac{1}{r^2}(1- e^{-2\lambda_0}) + 8\pi p_0+ \frac{1}{2}h_0 + (\rho_0+ p_0)h_{T0} . \label{20}
\end{eqnarray}
Besides, Eqs. (\ref{13}) and (\ref{18}) now assume the form
\begin{eqnarray}
&&\frac{dp_0}{dr} = -(\rho_0+ p_0)\psi'_0 ,\label{21}\\
&&\frac{2}{r}e^{-2\lambda_0}(\psi'_0 + \lambda'_0) = (8\pi + h_{T0})(\rho_0 + p_0) . \label{22}
\end{eqnarray}

Integrating Eq (\ref{19}), we have
\begin{equation}\label{23}
    e^{-2\lambda_0} = 1 - \frac{2m(r)}{r} ,
\end{equation}
where $m(r)$ is the total gravitational mass enclosed in the sphere of radius $r$, given by 
\begin{equation}\label{24}
  m(r) = 4\pi\int_0^r \bar{r}^2\rho_0(\bar{r})d\bar{r} - \frac{1}{4}\int_0^r \bar{r}^2 h_0(\bar{r}) d\bar{r} ,
\end{equation}
or alternatively $m = m_\rho + m_{\rm eff}$, where $m_\rho$ is the traditional mass as defined in GR and $m_{\rm eff}$ is an extra effective mass associated with the modification of Einstein's theory. The total mass of the compact star is given by $M \equiv m(r_{\rm sur}) $  where $r_{\rm sur}$ denotes the radial coordinate at the stellar surface where the pressure vanishes, i.e. $p(r= r_{\rm sur}) = 0$.

Taking into account Eq. (\ref{23}), Eq. (\ref{20}) reads
\begin{align}\label{25}
 \frac{d\psi_0}{dr} =& \left[ \frac{m}{r^2} + 4\pi rp_0 + \frac{r}{2}\left( \frac{1}{2}h_0 + (\rho_0+ p_0)h_{T0} \right) \right] \left( 1- \frac{2m}{r} \right)^{-1} ,
\end{align}
and therefore the relativistic structure of a compact star within the context of $f(R,T) = R+ h(T)$ gravity with conservative energy-momentum tensor is described by the following modified TOV equations
\begin{subequations}
\begin{align}
  \frac{dm}{dr} =&\ 4\pi r^2\rho - \frac{r^2}{4}h ,   \label{26a}  \\
  \frac{dp}{dr} =& -(\rho + p)\left[ \frac{m}{r^2} + 4\pi rp + \frac{r}{2}\left( \frac{1}{2}h + (\rho+ p)h_T \right) \right] \left( 1- \frac{2m}{r} \right)^{-1},  \label{26b}   \\
  \frac{d\psi}{dr} =& -\frac{1}{\rho+ p}\frac{dp}{dr} ,    \label{26c}
\end{align}
\end{subequations}
where we have removed the zero subscript because all quantities correspond to hydrostatic equilibrium. It is evident that when $h= 0$ one recovers the traditional TOV equations in the pure general relativistic case \cite{Tolman1939, Oppenheimer1939}. Given an EoS in the form $p = p(\rho)$, the system of differential equations (\ref{26a})-(\ref{26c}) can be solved by using the boundary conditions:
\begin{align}\label{27}
m(0) &= 0 ,   &   \rho(0) &= \rho_c ,   &   \psi(r_{\rm sur}) &= \frac{1}{2}\ln \left[ 1 - \frac{2M}{r_{\rm sur}} \right] .&
\end{align}

Notice that the exterior spacetime is still described by the Schwarzschild vacuum solution since $R= 0$ according to Eq. (\ref{8c}). It is interesting to note that the third term within the brackets on the right hand side of Eq. (\ref{26b}) could be interpreted as an extra force that modifies the standard stellar equilibrium in GR.

Furthermore, we must point out that the function $h(T)$ must satisfy Eq. (\ref{9}). In that regard, for the index $\nu =1$, one gets 
\begin{equation}\label{28}
(\rho + p)\frac{d}{dr}(\ln h_T) + \frac{1}{2}(\rho' + 3p') = 0 ,
\end{equation}
where we used $T= -\rho+ 3p$ in view of Eq. (\ref{11}). For a polytropic EoS given in the form $p = K\rho^\gamma$ with $K$ being some constant, the expression above becomes 
\begin{equation}\label{29}
(\rho + K\rho^\gamma)\frac{1}{h_T}\frac{dh_T}{d\rho} + \frac{1}{2}(1 + 3K\gamma\rho^{\gamma-1}) = 0 .
\end{equation}

Such equation can be integrated to give
\begin{equation}\label{30}
  h_T(\rho) = \alpha(\rho + K\rho^\gamma)^{\frac{1-3\gamma}{2\gamma-2}} \rho^{\frac{\gamma}{\gamma-1}} ,
\end{equation}
with $\alpha$ being an integration constant. We can also write 
\begin{equation}\label{31}
  h_T(\rho) \equiv \frac{dh}{dT} = \frac{dh}{d\rho}\frac{d\rho}{dT} = \frac{1}{-1 + 3K\gamma\rho^{\gamma-1}}\frac{dh}{d\rho} ,
\end{equation}
and from Eq. (\ref{30}), one obtains 
\begin{equation}\label{32}
  \frac{dh}{d\rho} = \alpha(3K\gamma\rho^{\gamma-1} - 1)(\rho + K\rho^\gamma)^{\frac{1-3\gamma}{2\gamma-2}} \rho^{\frac{\gamma}{\gamma-1}} .
\end{equation}

The solution of Eq. (\ref{32}) is given by 
\begin{align}\label{33}
h(\rho) =&\ \frac{2\alpha}{3\gamma-2}\rho^{\frac{\gamma}{\gamma-1}} \left[ \frac{K+ \rho^{1-\gamma}}{K} \right]^\frac{1-3\gamma}{2-2\gamma}(\rho+ K\rho^\gamma)^{\frac{1-3\gamma}{2\gamma-2}} \left\lbrace \rho\ _2F_1\left(\frac{1-3\gamma}{2-2\gamma}, \frac{2-3\gamma}{2-2\gamma}; \frac{4-5\gamma}{2-2\gamma}; -\frac{\rho^{1-\gamma}}{K}\right)  \right.   \nonumber   \\
&\left. - 3K(3\gamma-2)\rho^\gamma\ _2F_1\left(\frac{1-3\gamma}{2-2\gamma}, \frac{\gamma}{2\gamma-2}; 1+\frac{\gamma}{2\gamma-2}; -\frac{\rho^{1-\gamma}}{K}\right)  \right\rbrace , 
\end{align}
where $_2F_1(a,b;c;z)$ is the Gaussian hypergeometric function. The most general form has an extra integration constant and should be written as $h_G(\rho) = h(\rho) + \beta$, where $h(\rho)$ is the expression (\ref{33}) above. We note that one should pick $\beta=8\alpha/(3\sqrt{K})$ in order to get $h_G(\rho=0)=0 $. We did not do so because $\beta=0$ is the only choice that yields a fixed form of $h(T)$ (i.e., for all $K$ and $\gamma$) and hence a fixed extended theory of Gravity. Our choice for $\beta$ does correspond to a large negative effective cosmological constant  $\Lambda_{\rm eff}\sim 10^{-10}\rm m^{-2}$ (we recall the reader that the current observational value is $\Lambda_{\rm obs}\sim 10^{-56}\rm m^{-2}$ \cite{Aghanim2020}).

Due to the proper volume element $dV_{\rm pr} = \sqrt{^{(3)}g}d^3x = e^\lambda r^2\sin\theta drd\theta d\phi$, the total proper mass is given by \cite{Wald}
\begin{equation}\label{34}
M_{\rm pr} = \int \rho_{\rm rest}\sqrt{^{(3)}g}d^3x = 4\pi\int_0^{r_{\rm sur}} e^\lambda r^2\rho_{\rm rest}dr , 
\end{equation}
where $\rho_{\rm rest} \equiv m_{\rm N} n$ is the rest mass density, with $n$ being the number density of particles of mass $m_{\rm N}$.

We follow Ref. \cite{Bludman} and write the polytropic EoS $p = K\rho^\gamma$ in terms of the number density as
\begin{equation}
    \rho(n) = m_{\rm N}n \left[ 1- K(m_{\rm N}n)^{\gamma-1} \right]^\frac{1}{1- \gamma} ,
\end{equation}
or alternatively,
\begin{equation}\label{36}
    \rho_{\rm rest} = \frac{\rho}{ \left[ 1+ K\rho^{\gamma-1} \right]^\frac{1}{\gamma-1} } ,
\end{equation}
so that the total proper mass in Eq. (\ref{34}) becomes 
\begin{equation}\label{37}
    M_{\rm pr} = 4\pi\int_0^{r_{\rm sur}} \frac{ e^\lambda r^2\rho}{ \left[ 1+ K\rho^{\gamma-1} \right]^\frac{1}{\gamma-1} } dr .
\end{equation}

The difference between $M_\rho \equiv m_\rho(r_{\rm sur})$ and $M_{\rm pr}$ is known as the \textit{gravitational binding energy} of the stellar configuration, namely
\begin{equation}\label{38}
   E_B = M_\rho - M_{\rm pr} .
\end{equation}
This binding energy will be plotted as a function of the compact star proper mass in order to analyse the stellar stability of the solutions. Just as in the pure GR case, a cusp must be observed when $E_B$ is minimum in any other theory of modified gravity. In that respect, we expect that such a cusp indicates the instability of compact stars beyond the minimum-binding-energy point in $f(R,T)=R+ h(T)$ gravity.


\section{Pulsation equations}\label{Sect3}

To address the problem of radial stability of compact stars in any theory of gravity, we have to calculate the vibration frequencies of the normal modes. This means that a fluid element located at $r$ in the hydrostatic equilibrium is displaced to the radial coordinate $r+ \xi(t,r)$ in the perturbed system. Moreover, the small radial oscillations can be written as $F(t,r) = F_0(r)+ \delta F(t,r)$ where $F$ represents any metric or thermodynamic variable and $\delta F$ is the Eulerian perturbation. Following Ref. \cite{Chandrasekhar}, it is convenient to define $v\equiv dr/dt = \partial\xi/\partial t$ so that the four-velocity can be given by $u^\mu = (e^{-\psi}, e^{-\psi_0}v, 0, 0)$ in the linear approximation. Thus, the energy-momentum tensor (\ref{11}) to first order takes the following form 
\begin{equation}
T_\mu^{\ \nu}=  \begin{pmatrix}
  -\rho & -(\rho_0 + p_0)v & 0 & 0 \\
  (\rho_0+p_0)ve^{2\lambda_0- 2\psi_0} & p & 0 & 0 \\
  0 & 0 & p & 0 \\
  0 & 0 & 0 & p
\end{pmatrix} .
\end{equation}

Henceforth we only preserve linear terms in the perturbations. In view of Eq. (\ref{19}), Eq. (\ref{14}) reads
\begin{equation}\label{40}
 2\frac{\partial}{\partial r}(re^{-2\lambda_0}\delta\lambda) = \left( 8\pi - \frac{1}{2}\frac{dh}{d\rho}\bigg\vert_{\rho_0} \right)r^2\delta\rho .
\end{equation}
Besides, using Eq. (\ref{20}), from Eq. (\ref{15}) we obtain 
\begin{align}\label{41}
  \frac{2}{re^{2\lambda_0}}\left[ \frac{\partial}{\partial r}(\delta\psi) - 2\psi'_0\delta\lambda \right] =&\ \frac{2}{r^2}e^{-2\lambda_0}\delta\lambda + (8\pi + h_{T0})\delta p  \nonumber   \\
  &+ \left[ h_{T0}+ \frac{1}{2}\frac{dh}{d\rho}\bigg\vert_{\rho_0} + (\rho_0+ p_0)\frac{dh_T}{d\rho}\bigg\vert_{\rho_0} \right]\delta\rho , 
\end{align}
and Eq. (\ref{17}) assumes the form 
\begin{equation}\label{42}
  \frac{2}{r}e^{-2\lambda_0}\delta\lambda = -(8\pi + h_{T0})(\rho_0 + p_0)\xi ,
\end{equation}
or alternatively, using Eq. (\ref{22}):
\begin{equation}\label{43}
 \delta\lambda = -(\psi'_0+ \lambda'_0)\xi .
\end{equation}

Given Eq. (\ref{21}), the linearized form of Eq. (\ref{13}) leads to
\begin{equation}\label{44}
  (\rho_0+ p_0)e^{2\lambda_0- 2\psi_0}\frac{\partial v}{\partial x^0} + \frac{\partial}{\partial r}(\delta p) + (\rho_0+ p_0)\frac{\partial}{\partial r}(\delta\psi)  + (\delta\rho + \delta p)\psi'_0 = 0 , 
\end{equation}
and taking into account Eq. (\ref{42}), Eq. (\ref{40}) reads
\begin{align}
  \delta\rho &= -\frac{\rho'_0}{\mathcal{B}_1}\frac{dh_T}{d\rho}\bigg\vert_{\rho_0} (\epsilon_0+ p_0)\xi - \frac{8\pi + h_{T0}}{r^2 \mathcal{B}_1} \frac{\partial}{\partial r}\left[ (\rho_0+ p_0)r^2\xi \right]   \nonumber \\
  &= -\frac{\mathcal{B}_2}{\mathcal{B}_1}\rho'_0\xi - \frac{8\pi + h_{T0}}{\mathcal{B}_1}(\rho_0 + p_0) \frac{e^{\psi_0}}{r^2}\frac{\partial}{\partial r}(r^2\xi e^{-\psi_0}) ,  \label{45}
\end{align}
where we have defined 
\begin{eqnarray}
  \mathcal{B}_1 &\equiv& 8\pi - \frac{1}{2}\frac{dh}{d\rho}\bigg\vert_{\rho_0} ,  \label{46}  \\
  \mathcal{B}_2 &\equiv& 8\pi + h_{T0} + (\rho_0 + p_0)\frac{dh_T}{d\rho}\bigg\vert_{\rho_0} .  \label{47}
\end{eqnarray}

By means of Eqs. (\ref{22}) and (\ref{42}), Eq. (\ref{41}) can also be written as
\begin{equation}\label{48}
   (\rho_0 + p_0)\frac{\partial}{\partial r}(\delta\psi) = \left[ \delta p + \frac{\mathcal{B}_2- \mathcal{B}_1}{8\pi + h_{T0}}\delta\rho  - (\rho_0 + p_0)\left( 2\psi'_0 + \frac{1}{r} \right)\xi \right](\psi'_0 + \lambda'_0) .
\end{equation}

Let us now suppose that all perturbations have a harmonic time dependence, this is, $\xi(t,r) = \chi(r)e^{i\omega t}$ and $\delta F(t,r)= \delta F(r)e^{i\omega t}$, where $\omega$ is the frequency of radial pulsations to be determined, and the amplitudes of the perturbations depend only on the radial coordinate. Such an assumption will allow us to obtain time-independent equations since the exponential factor cancels out in each perturbed variable. As a consequence, all terms are now amplitudes of the perturbations and quantities of the static background, and hence we can remove the subscript zero. According to Eqs. (\ref{29})-(\ref{32}), we get $\mathcal{B}_1= \mathcal{B}_2$. In view of Eq. (\ref{48}), Eq. (\ref{44}) reads 
\begin{equation}\label{49}
  \omega^2(\rho+ p)e^{2\lambda - 2\psi}\chi = \frac{d}{dr}(\delta p) + (2\psi' + \lambda')\delta p + \psi'\delta\rho  - (\rho + p)\left( 2\psi' + \frac{1}{r} \right)(\psi' + \lambda')\chi . 
\end{equation}

For a barotropic EoS $p= p(\rho)$, we can express the Eulerian perturbation of the pressure as $\delta p= (dp/d\rho)\delta\rho$. So through Eq. (\ref{45}), we have
\begin{equation}\label{50}
  \delta p = -\frac{dp}{d\rho}\rho'\chi - \frac{8\pi + h_{T}}{\mathcal{B}_1}\frac{dp}{d\rho}(\rho + p) \frac{e^{\psi}}{r^2}\frac{d}{d r}(r^2\chi e^{-\psi}) ,
\end{equation}
or alternatively,
\begin{equation}\label{51}
\delta p = -\chi p' - \frac{8\pi + h_{T}}{\mathcal{B}_1}\Gamma p \frac{e^\psi}{r^2}\frac{d}{dr}(r^2\chi e^{-\psi}) ,
\end{equation}
where $\Gamma = (1+ \rho/p)dp/d\rho$ is the adiabatic index at constant entropy. Then the Lagrangian perturbation of the pressure is given by 
\begin{equation}\label{52}
 \Delta p \equiv \delta p + \chi p' = - \frac{8\pi + h_{T}}{\mathcal{B}_1}\Gamma p \frac{e^\psi}{r^2}\frac{d}{dr}(r^2\chi e^{-\psi}) . 
\end{equation}

In addition, Eq. (\ref{45}) can be written as 
\begin{equation}\label{53}
     \delta\rho = -\frac{\rho'}{\mathcal{B}_1}\frac{dh_T}{d\rho}(\rho+ p)\chi - \frac{8\pi + h_{T}}{\mathcal{B}_1}\mathcal{Q} , 
\end{equation}
where
\begin{equation}\label{54}
     \mathcal{Q} \equiv \frac{d}{dr}[(\rho + p)\chi] + \frac{2}{r}(\rho + p)\chi .
\end{equation}

Thus, by means of Eqs. (\ref{52}) and (\ref{53}) together with the $\theta\theta$-component of the field equations (\ref{16}) in the state of hydrostatic equilibrium, the expression (\ref{49}) gives
\begin{align}\label{55}
    \omega^2(\rho + p)e^{2\lambda- 2\psi}\chi =&\ (\Delta p)' + (2\psi'+ \lambda') \Delta p + \frac{4}{r}\chi p' + \chi(\rho + p) \left[ 8\pi p + \frac{1}{2}h+ (\rho + p)h_T \right]e^{2\lambda}   \nonumber  \\
    &- \chi(\rho +p)\psi'^2 + \left[ 1- \frac{8\pi + h_{T}}{\mathcal{B}_1} \right]\psi'\mathcal{Q} - \frac{\rho'}{\mathcal{B}_1}\frac{dh_T}{d\rho}(\rho + p)\psi'\chi , 
\end{align}
and using Eq. (\ref{52}), $\mathcal{Q}$ can be given by 
\begin{equation}\label{56}
    \mathcal{Q} = \rho'\chi - \frac{\mathcal{B}_1}{8\pi +h_T}\frac{\rho + p}{\Gamma p}\Delta p .
\end{equation}

Finally, Eq. (\ref{55}) becomes 
\begin{align}\label{57}
   \frac{d}{dr}(\Delta p) =&\ \chi \left\lbrace  \omega^2(\rho + p)e^{2\lambda- 2\psi} - (\rho + p) \left[ 8\pi p + \frac{1}{2}h+ (\rho + p)h_T \right]e^{2\lambda} + (\rho +p)\psi'^2 - \frac{4}{r} p'  \right.  \nonumber  \\
   &\hspace{-1cm}\left. + \frac{\psi'\rho'}{\mathcal{B}_1}\left[ 8\pi +h_T -\mathcal{B}_1 + (\rho + p)\frac{dh_T}{d\rho} \right] \right\rbrace - \Delta p \left[ 2\psi'+ \lambda' + \left( 1- \frac{\mathcal{B}_1}{8\pi + h_T} \right)\frac{\rho + p}{\Gamma p}\psi' \right] .
\end{align}

If we define $\zeta \equiv \chi/r$, from Eqs. (\ref{52}) and (\ref{57}) we obtain the first-order differential equations governing the radial oscillations for a perfect fluid sphere in $f(R,T) = R+ h(T)$ gravity with conservative energy-momentum tensor, namely
\begin{align}
  \frac{d\zeta}{dr} =& -\frac{1}{r}\left[ 3\zeta + \frac{\mathcal{B}_1}{8\pi +h_T}\frac{\Delta p}{\Gamma p} \right] + \psi'\zeta ,   \label{58}  \\
  \frac{d}{dr}(\Delta p) =&\ \zeta \left\lbrace  \omega^2(\rho + p)e^{2\lambda- 2\psi}r - (\rho + p) \left[ 8\pi p + \frac{1}{2}h+ (\rho + p)h_T \right]re^{2\lambda} - 4p'  \right.  \nonumber  \\
  &\left. + (\rho +p)r\psi'^2  + \frac{r}{\mathcal{B}_1}\left[ 8\pi + h_T -\mathcal{B}_1 + (\rho + p)\frac{dh_T}{d\rho} \right]\psi'\rho' \right\rbrace  \nonumber  \\ 
   &- \Delta p \left[ 2\psi'+ \lambda' + \left( 1- \frac{\mathcal{B}_1}{8\pi + h_T} \right)\frac{\rho + p}{\Gamma p}\psi' \right] .  \label{59}
\end{align}

It is evident that when $h = 0$ (which implies $h_T =0$ and $\mathcal{B}_1= 8\pi$), we retrieve the radial pulsation equations corresponding to the pure general relativistic case \cite{Gondek, Flores2010, Pereira2018, Pretel2020MNRAS}. Here we should note that Eq. (\ref{58}) has a singularity at the stellar origin ($r= 0$). Therefore, in order that $d\zeta/dr$ be finite at any spacetime point inside the star, it is required that as $r \rightarrow 0$ the coefficient of $1/r$ term must vanish, so 
\begin{equation}\label{60}
    \Delta p = -3\Gamma p\zeta \left[\frac{8\pi + h_T}{\mathcal{B}_1}\right]  \qquad \  \text{as}  \qquad  \  r\rightarrow 0 ,
\end{equation}
and since the surface of the star is determined by the condition $p(r= r_{\rm sur})=0$, we must demand the following boundary condition for the Lagrangian perturbation of the pressure
\begin{equation}\label{61}
    \Delta p = 0  \qquad \  \text{as}  \qquad  \  r\rightarrow r_{\rm sur} .
\end{equation}


\section{Equation of state}\label{Sect4}

Since the polytropic EoS is analytically the simplest for the study of compact stars, here we will use such two equations of state widely adopted in the literature to describe neutron stars:
\begin{itemize}
\item[-] EoS I: $p= K_1\rho^{\gamma_1}$ where $\gamma_1= 2$ and $K_1= 10^8\ \rm m^2$ \cite{KokkotasR2001, Allen1998}. According to Eqs. (\ref{30}) and (\ref{33}), we obtain 
\begin{equation}\label{62}
    h_T(\rho) = \frac{\alpha_1}{\rho^3} \left( K_1 + \frac{1}{\rho} \right)^{-5/2} ,
\end{equation}
and
\begin{equation}\label{63}
    h(\rho) = -\frac{2\alpha_1}{3}\frac{\sqrt{\rho}}{(1+K_1\rho)^{3/2}}  \left[ 3+ 4\sqrt{1+ \frac{1}{K_1\rho}}  + 4K_1\rho\left( \sqrt{1+ \frac{1}{K_1\rho}} -1 \right) \right] . 
\end{equation}
The total proper mass (\ref{37}) takes the form 
\begin{equation}\label{64}
    M_{\rm pr} = 4\pi\int_0^{r_{\rm sur}} \frac{ e^\lambda r^2\rho}{ 1+ K_1\rho } dr .
\end{equation}

\item[-] EoS II: $p= K_2\rho^{\gamma_2}$ where $\gamma_2= 5/3$ and $K_2= 1.2237 \times 10^5\ \rm m^{4/3}$ ($= 1.475 \times 10^{-3}\ [\rm fm^3/MeV]^{2/3}$ in physical units) \cite{PhysRevD.88.084023, Moraes2016}. As a result, Eq. (\ref{30}) becomes 
\begin{equation}\label{65}
h_T(\rho) = \frac{\alpha_2}{\rho^{5/2}} \left( K_2 + \frac{1}{\rho^{2/3}} \right)^{-3} ,
\end{equation}
and Eq. (\ref{33}) reduces to 
\begin{align}\label{66}
   h(\rho) =&\ \frac{3\alpha_2}{16} \left\lbrace \frac{2(-7+ 5K_2\rho^{2/3})\sqrt{\rho}}{(1+ K_2\rho^{2/3})^2}  + 5\left( \frac{-1}{K_2} \right)^{3/4}{\rm ArcTan}\left[ \left( \frac{-1}{K_2\rho^{2/3}} \right)^{1/4} \right]  \right.  \nonumber  \\
   &\left. + 5\left( \frac{-1}{K_2} \right)^{3/4}{\rm ArcTanh}\left[ \left( \frac{-1}{K_2\rho^{2/3}} \right)^{1/4} \right] \right\rbrace .
\end{align}

In this case, the total proper mass is given by 
\begin{equation}\label{67}
    M_{\rm pr} = 4\pi\int_0^{r_{\rm sur}} \frac{ e^\lambda r^2\rho}{ [1+ K_2\rho^{2/3}]^{3/2} } dr .
\end{equation}

\end{itemize}

Note that both free parameters $\alpha_1$ and $\alpha_2$ have the same units, i.e. $[\alpha_1]= [\alpha_2]= \rm m^{-1}$, so from now on we are going to refer to a single parameter $\alpha$. In addition, we will consider values of $\alpha$ for which an appreciable change is observed in the mass-radius relations.


\section{Numerical results and discussion} \label{Sect5}

\subsection{Equilibrium configurations}

\begin{figure}
\centering
 \includegraphics[width=4.56cm]{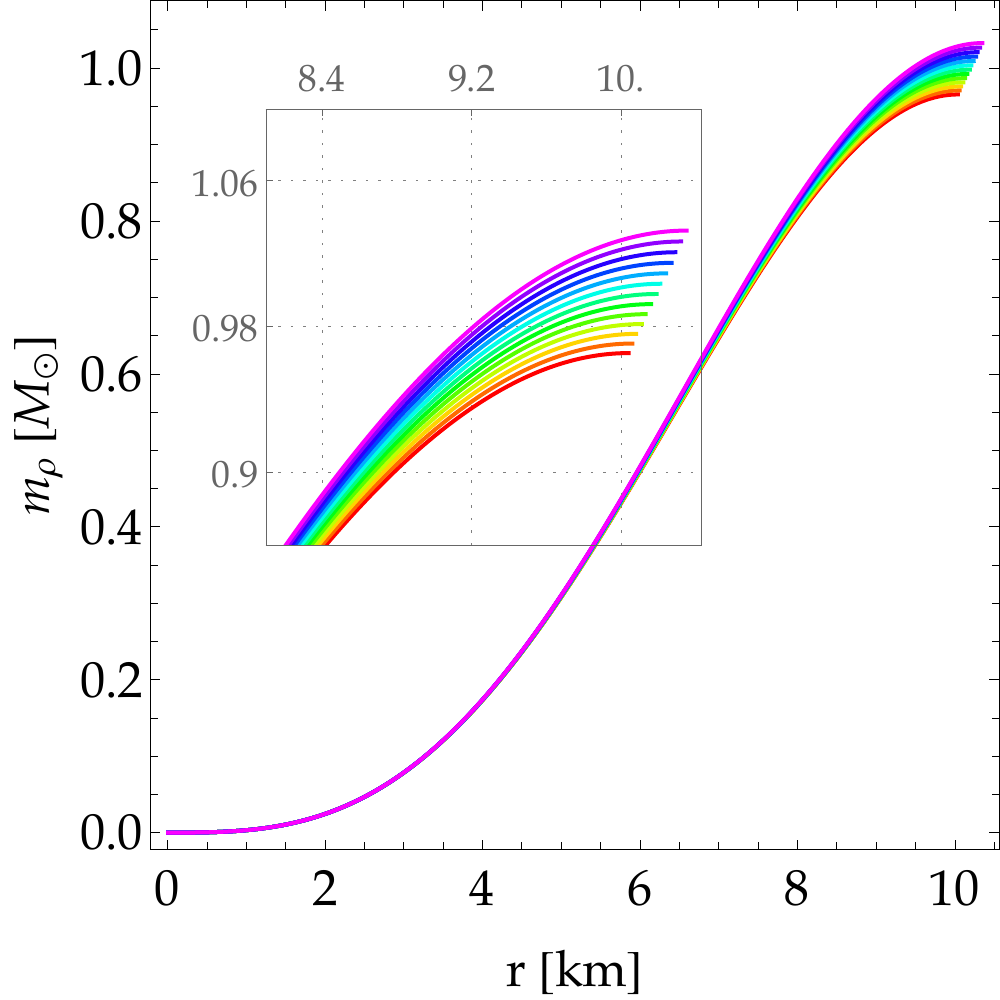}
 \includegraphics[width=4.85cm]{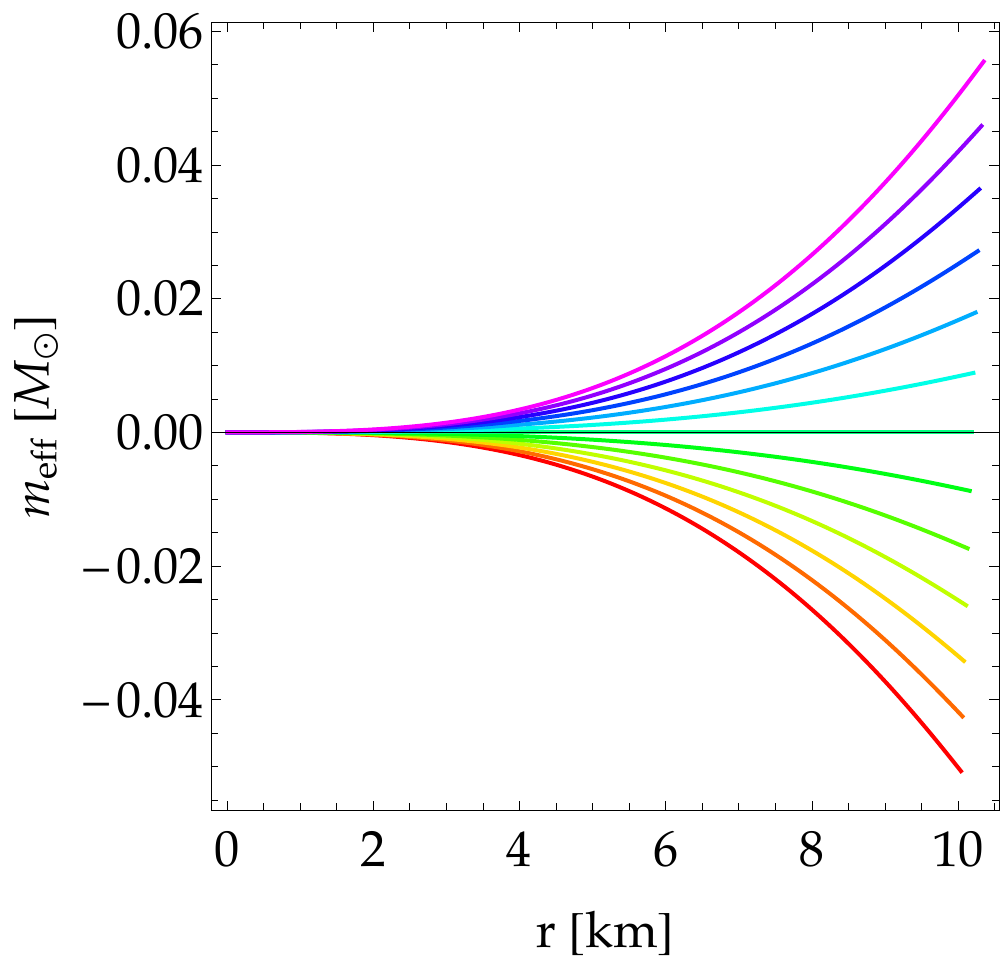}
 \includegraphics[width=5.79cm]{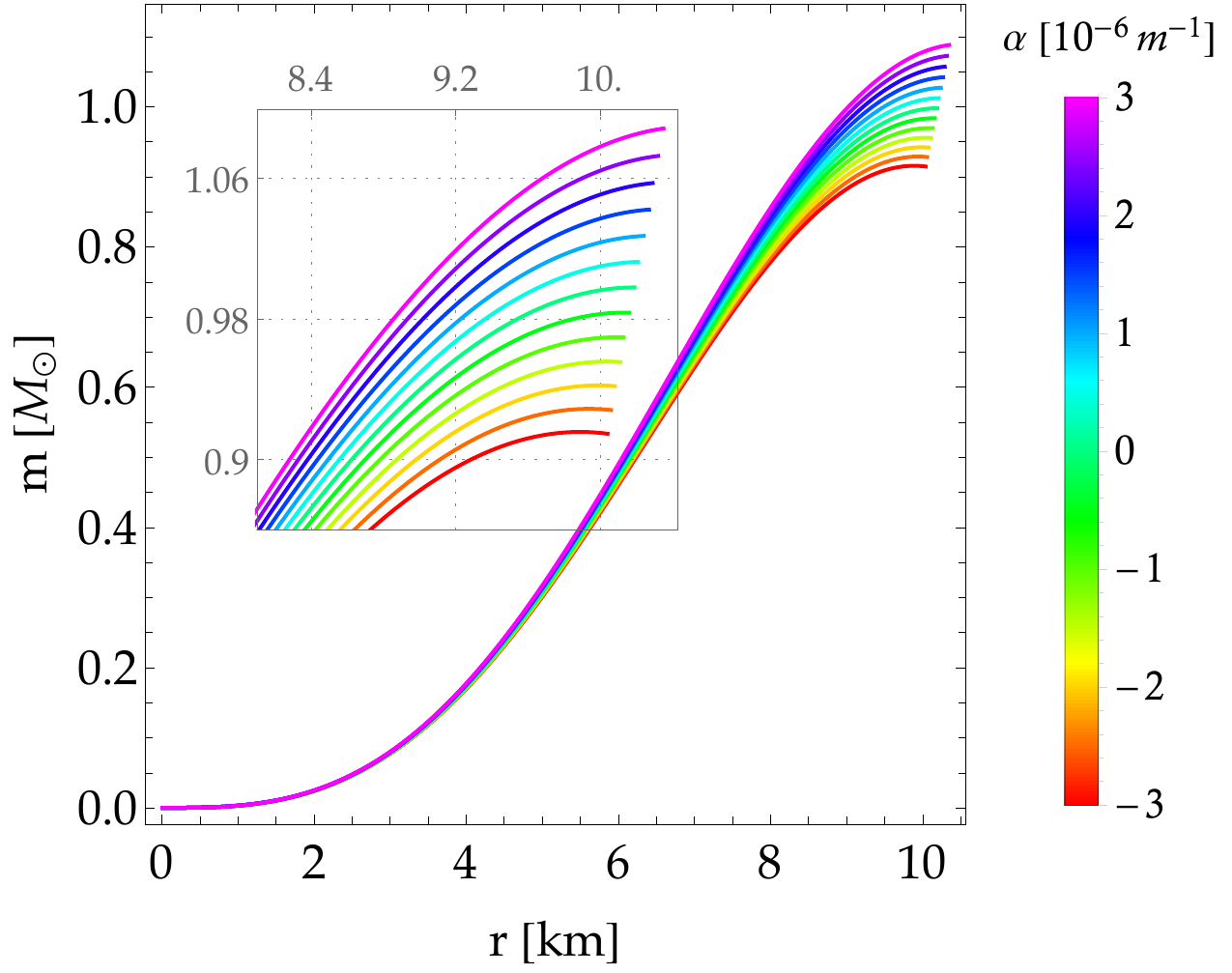}
 \caption{\label{figure1} On the left panel is shown the mass function of standard matter (\ref{StanMass}), on the middle panel the mass distribution associated with the $h(T)$ term which is calculated from Eq. (\ref{EffMass}), and on the right panel it is displayed the total gravitational mass of the star. These plots correspond to a central mass density $\rho_c = 1.5 \times 10^{18}\ \rm kg/m^3$ with polytropic EoS I for different values of the coupling constant $\alpha$ (see color scale on the right). We can observe that the deviations from GR occur mainly in the outer layers of the star, that is, the changes are more substantial as we approach the surface. }  
\end{figure}

Given the polytropic EoS I together with a specific value of the parameter $\alpha$ in Eqs. (\ref{62}) and (\ref{63}), the system of modified TOV equations (\ref{26a})-(\ref{26c}) with boundary conditions (\ref{27}) is numerically integrated from the stellar center at $r=0$ up to the surface $r= r_{\rm sur}$ where the fluid pressure vanishes. In particular, for a central mass density $\rho_c = 1.5 \times 10^{18}\ \rm kg/m^3$ and different values of $\alpha$, we obtain the solutions shown in Fig. \ref{figure1}. Once the mass density $\rho(r)$ is determined from the stellar structure equations, we can calculate the mass distribution of standard matter and the mass function associated with the effective fluid through the following integrals (see also Eq. (\ref{24})), respectively,
\begin{eqnarray}
  m_\rho(r) &=& 4\pi\int_0^r \bar{r}^2\rho(\bar{r})d\bar{r} ,  \label{StanMass}   \\
  m_{\rm eff}(r) &=& - \frac{1}{4}\int_0^r \bar{r}^2 h(\bar{r}) d\bar{r} ,  \label{EffMass}
\end{eqnarray}
so that at the stellar surface, we have $M= M_\rho + M_{\rm eff}$. According to the data recorded in Table \ref{table1}, we observe that for a given $\rho_c$, as $\alpha$ increases the surface radius $r_{\rm sur}$ and total mass $M$ of the star also increase. The standard mass distribution (\ref{StanMass}) is shown on the left plot of Fig. \ref{figure1} for several values of $\alpha$\footnote{This free parameter is given in geometric units throughout the text, so we consider $[\alpha]= \rm m^{-1}$ in all figures.}. The intermediate plot of the same figure indicates that for positive (negative) values of $\alpha$ the effective mass function associated with the $h(T)$ term is positive (negative). As a consequence, the gravitational mass $m(r)$ overcomes the GR counterpart for positive values of the coupling constant $\alpha$, while the opposite occurs for negative values of $\alpha$. This means that the modification of GR via the $h(T)$ term can admit characteristics that would normally be considered unphysical for an ordinary matter fluid. Furthermore, it is evident that the contribution of these effective masses is zero at the GR limit, this is, when $\alpha \rightarrow 0$. Nevertheless, we point out that even the maximal mass deviation from GR are below the current observational limit.

\begin{figure}
\centering
 \includegraphics[width=7.6cm]{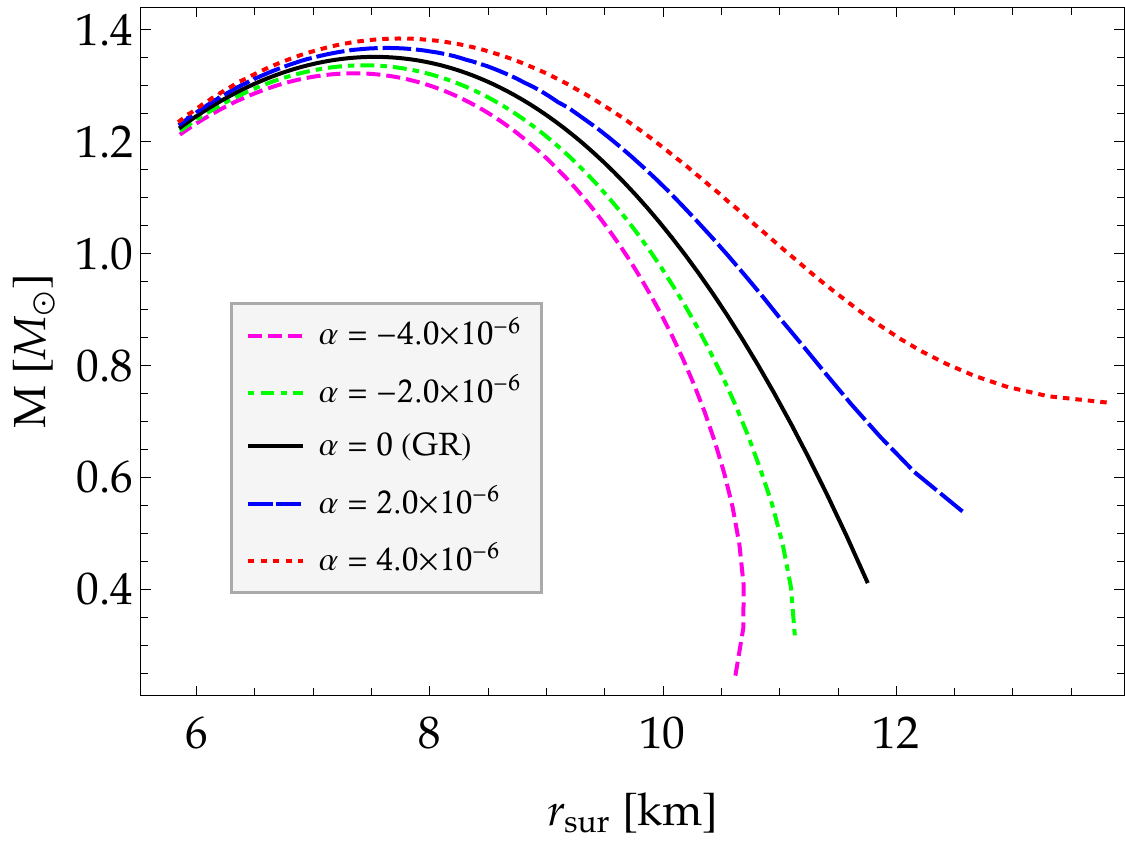}
 \includegraphics[width=7.6cm]{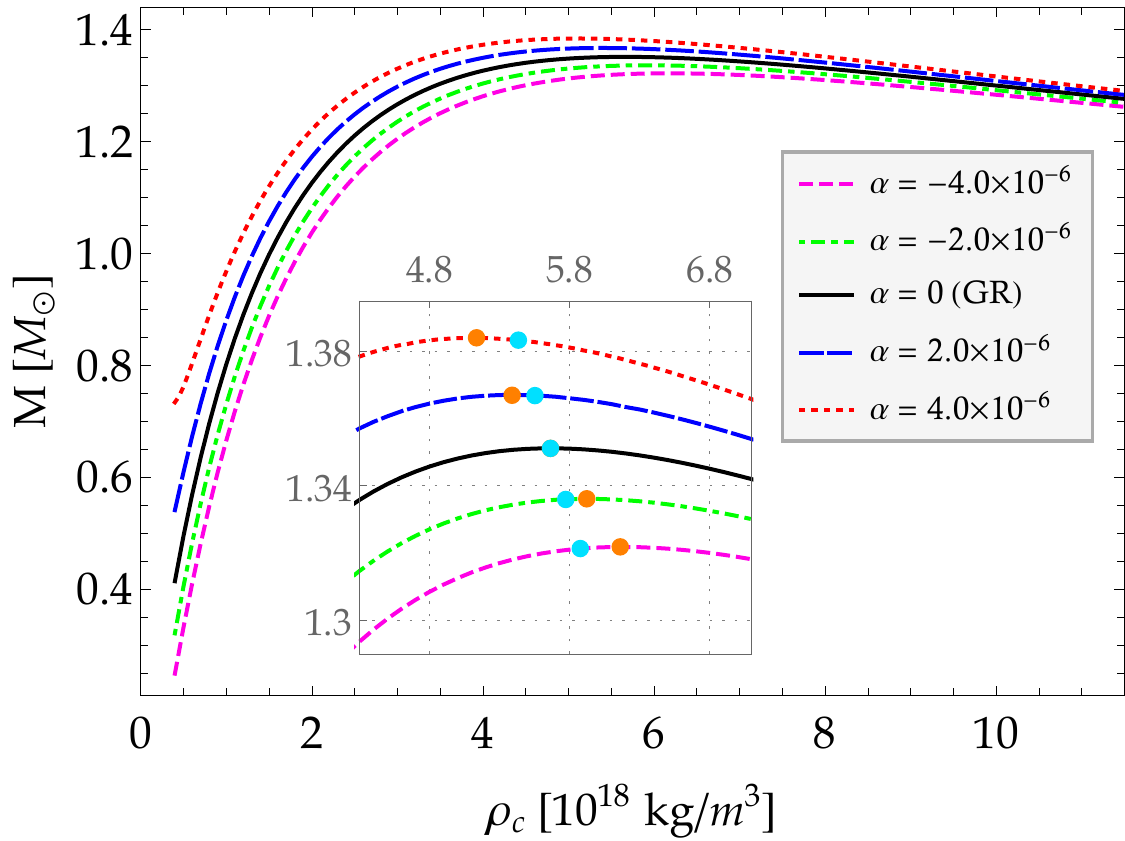}
 \caption{\label{figure2} Left panel: Mass-radius diagram for some values of $\alpha$ (given in geometric units as $\rm m^{-1}$), where $M$ represents the total gravitational mass at the surface of the star. Right panel: Mass-central density relation, where the central density corresponds to the standard matter. The GR solution is shown in all plots as a benchmark by a solid black line. The more pronounced deviations for $r_{\rm sur}$ from GR ($\alpha=0$) take place at low central densities, whereas for large masses (close to the maximum mass) the changes are very slight due to the $h(T)$ term. Moreover, the full orange and cyan circles indicate the maximum-mass points and the minimum-binding-energy points, respectively. See also the right plot of Fig. \ref{figure4} to identify the minimum-binding energy. }  
\end{figure}

\begin{figure}
\centering
 \includegraphics[width=7.8cm]{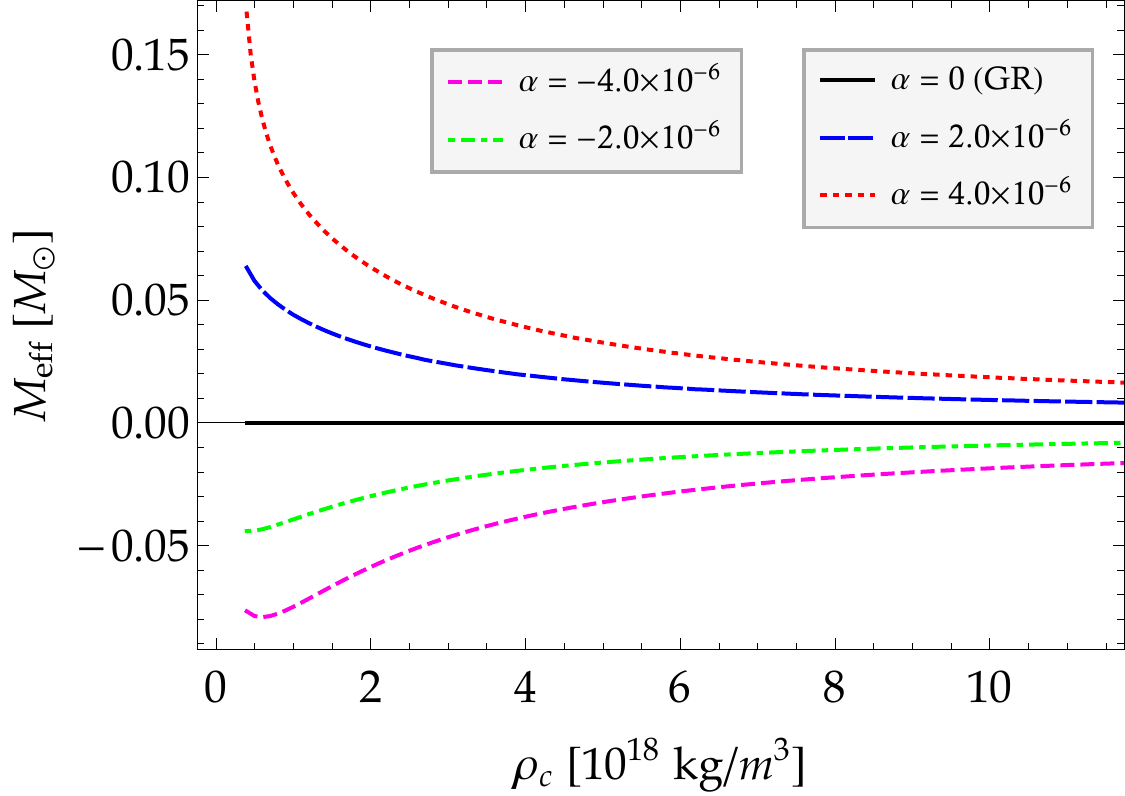}
 \caption{\label{figure3} Effective mass at $r=r_{\rm sur}$ as a function of the central density. Different styles and colors of the curves correspond to different values of the parameter $\alpha$, and it is evident that at the GR limit there is no effective mass. Furthermore, we can see that $\vert M_{\rm eff} \vert$ assumes higher values in the low-central-densitiy-region and which explains the behavior in the plots of Fig. \ref{figure2}. }  
\end{figure}

Using the central mass density as a parameter, the families of isotropic neutron stars in $f(R,T)= R+ h(T)$ gravity are presented in Fig. \ref{figure2} for several values of $\alpha$. According to the left panel, the total gravitational mass undergoes significant modifications as we move away from GR, mainly in the low-central-density region. One can also notice the qualitative change in the curve for $\alpha>0$ --- namely, the change in the sign of the concavity at large radii. For even larger values of $\alpha$ (not shown), the mass of the star and its radius will be greater than those observed, which yields a constraint on the maximum value of $\alpha$. The right plot of the same figure illustrates the behavior of the total mass as a function of the central mass density, where we have indicated the maximum-mass points by full orange circles. Figure \ref{figure3} shows the effective mass calculated at the radius of the star, this is, $M_{\rm eff} \equiv m_{\rm eff}(r_{\rm sur})$. For low enough central densities, the contributions from the $h(T)$ term are relatively significant compared to the higher-central-density region. In fact, this behavior explains the results obtained in Fig. \ref{figure2}.

\begin{figure}
\centering
 \includegraphics[width=7.6cm]{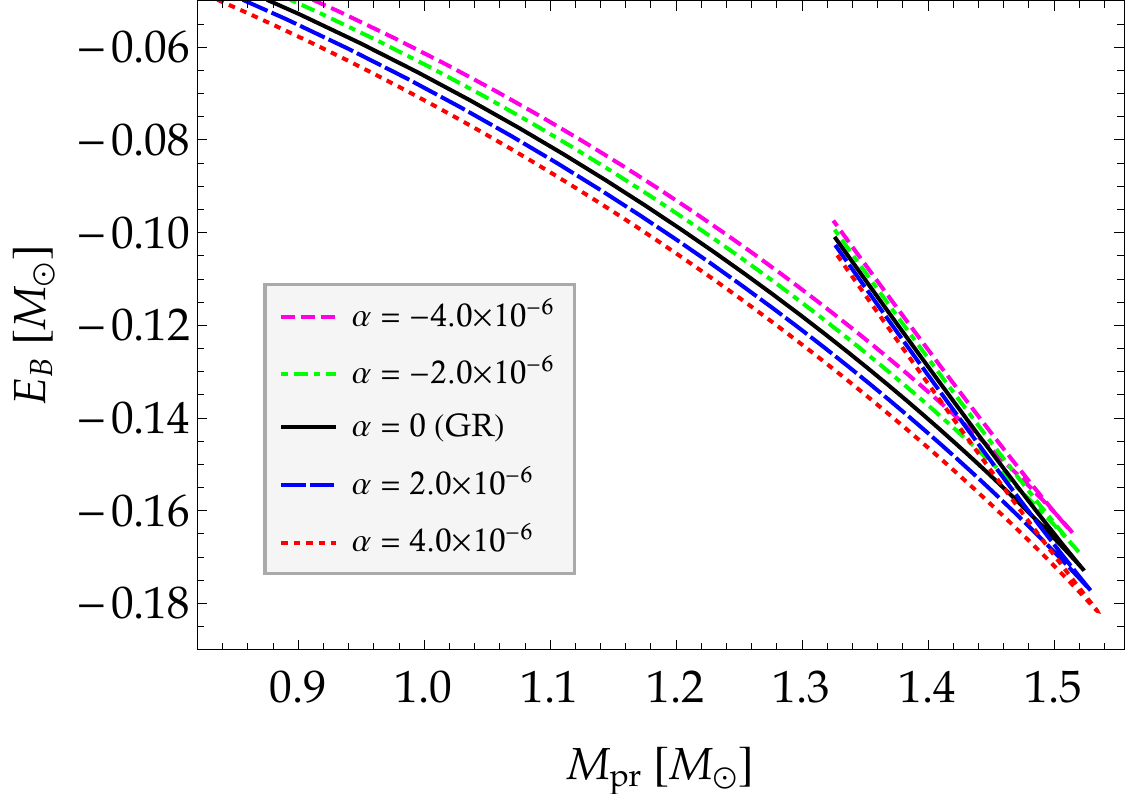}
 \includegraphics[width=7.6cm]{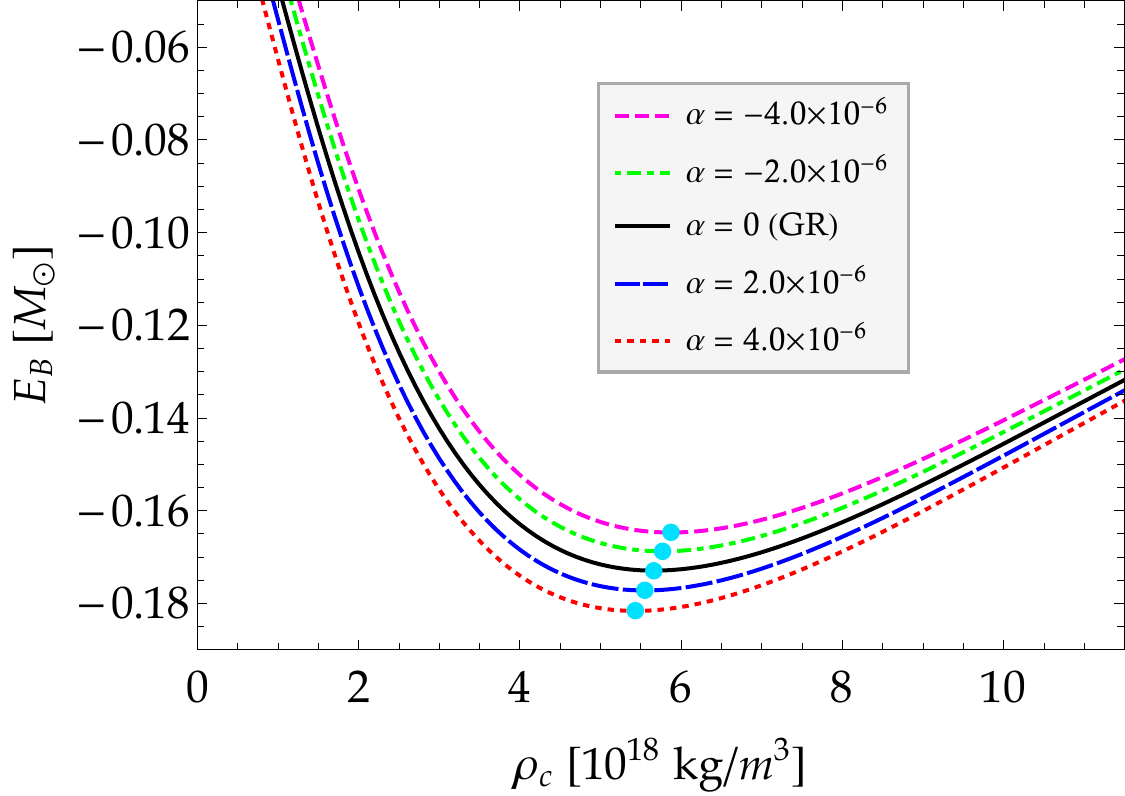}
 \caption{\label{figure4} Gravitational binding energy as a function of the proper mass (left panel) and of the central mass density (right panel) for the stellar equilibrium configurations displayed in Fig. \ref{figure2}. We can see the formation of a cusp when the binding energy becomes minimal. Furthermore, the full cyan circles on the right panel indicate the minimum-binding-energy points where the central density values correspond to $\omega_0^2= 0$ as shown on the left plot of Fig. \ref{figure8}.  }  
\end{figure}

The stellar configurations presented in Fig. \ref{figure2} describe neutron stars in hydrostatic equilibrium. Such equilibrium can be either stable or unstable with respect to a small radial perturbation. It has been shown, at least in GR (see for instance Refs. \cite{Glendenning, Haensel2007}), that a turning point from stability to instability occurs when $dM(\rho_c)/d\rho_c = 0$. According to this criterion, the stable branch in the sequence of stars should be located below the critical central density corresponding to the maximum mass. Nevertheless, this must be examined more carefully in the present scenario because there is now an effective mass contribution that ends up altering the total mass of the stars. An indicator that could also establish the onset of instability is the formation of a cusp when the binding energy is a minimum. From the left plot of Fig. \ref{figure4}, we can appreciate that such a cusp arises for different values of the parameter $\alpha$ when we analyze the behavior of the binding energy (\ref{38}) in terms of the proper mass (\ref{64}). In addition, given a particular value of $\alpha$, the right plot of the same figure manifests a minimum for the binding energy corresponding to a certain value of central mass density indicated by full cyan circles.

Although in Einstein's gravity the maximum-mass point and the  minimum-binding-energy point coincide at the same value of central mass density, there is a small discrepancy when $\alpha \neq 0$ which can be visualized in the right plot of Fig. \ref{figure2}. Therefore, to have absolute certainty about the central density value from which the unstable branch begins, we will analyze the stellar stability with respect to adiabatic radial oscillations in the next subsection.

\begin{figure}
\centering
 \includegraphics[width=7.8cm]{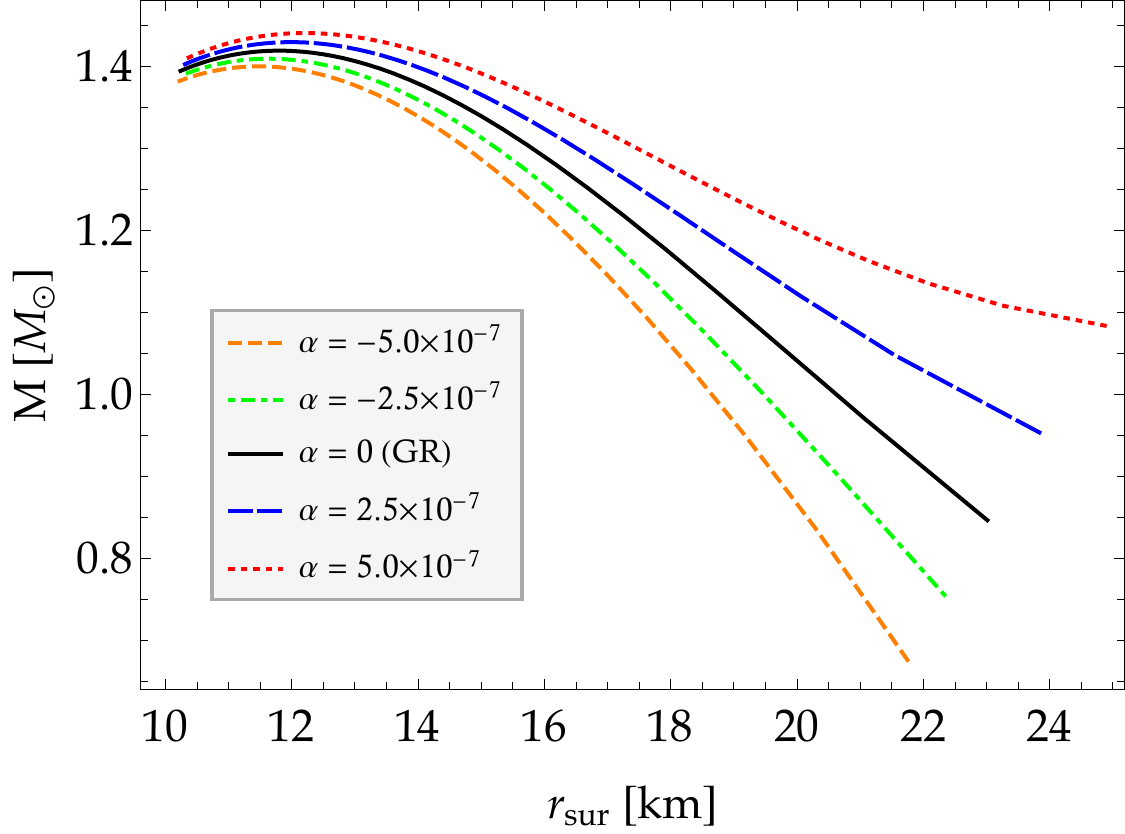}
 \caption{\label{figure5} Mass-radius relation for neutron stars with polytropic EoS II in the context of $f(R,T)= R+ h(T)$ gravity for several values of $\alpha$, where $h_T(\rho)$ and $h(\rho)$ are given by Eqs. (\ref{65}) and (\ref{66}), respectively. }  
\end{figure}

For neutron stars with polytropic EoS II, the mass-radius relations are shown in Fig. \ref{figure5}. Although the values for $\vert\alpha\vert$ are smaller than those considered in EoS I, the qualitative behavior in macroscopic quantities is similar for both equations of state. Nonetheless, here we must emphasize that our results differ considerably from those obtained by the authors in Ref. \cite{Santos2019}. In particular, they showed that the most relevant changes due to the $h(T)$ term occur for large masses, and obtained mass-radius curves that intersect at some value of central mass density. Our analysis shows on the contrary that the most pronounced modifications take place in the low masses region, and also the $M$-$r_{\rm sur}$ curves never intersect.

\begin{table}[t]
\centering
\begin{tabular}{|c|c|c|c|c|c|c|c|}
\hline
$\alpha$  &  $r_{\rm sur}$  &  $M_\rho$  &  $M_{\rm eff}$  &  $M$  &  $f_0$  &  $f_1$  &  $f_2$   \\
$[10^{-6} \rm m^{-1}]$  &  $[\rm{km}]$  &  $[M_\odot]$  &  $[M_\odot]$  &  $[M_\odot]$  &  [kHz]  &  [kHz]  &  [kHz]  \\
\hline
$-4.0$  &  9.993  &  0.955  &  $-0.067$  &  0.888  &  2.580  &  6.145  &  9.069  \\
$-2.0$  &  10.088  &  0.976  &  $-0.034$  &  0.942  &  2.443  &  5.943  &  8.792  \\
$0$ (GR)  &  10.190  &  0.998  &  0  &  0.998  &  2.301  &  5.734  &  8.506  \\
$2.0$  &  10.296  &  1.021  &  0.036  &  1.057  &  2.152  &  5.517  &  8.208  \\
$4.0$  &  10.408  &  1.045  &  0.075  &  1.120  &  1.997  &  5.292  &  7.899  \\
\hline
\end{tabular}
\caption{\label{table1} 
Neutron star properties with central mass density $\rho_c = 1.5 \times 10^{18}\ \rm{kg}/\rm{m}^3$ and EoS I in $f(R,T)= R+ h(T)$ gravity for different values of the coupling constant $\alpha$ (given in geometric units). The radial behavior of the standard, effective and total gravitational mass function of these stars is shown in Fig. \ref{figure1}. In addition, we present the pulsation frequencies $f_n= \omega_n/2\pi$ for the first three normal modes.}
\end{table}

\subsection{Radial pulsations}

\begin{figure}
\centering
 \includegraphics[width=7.68cm]{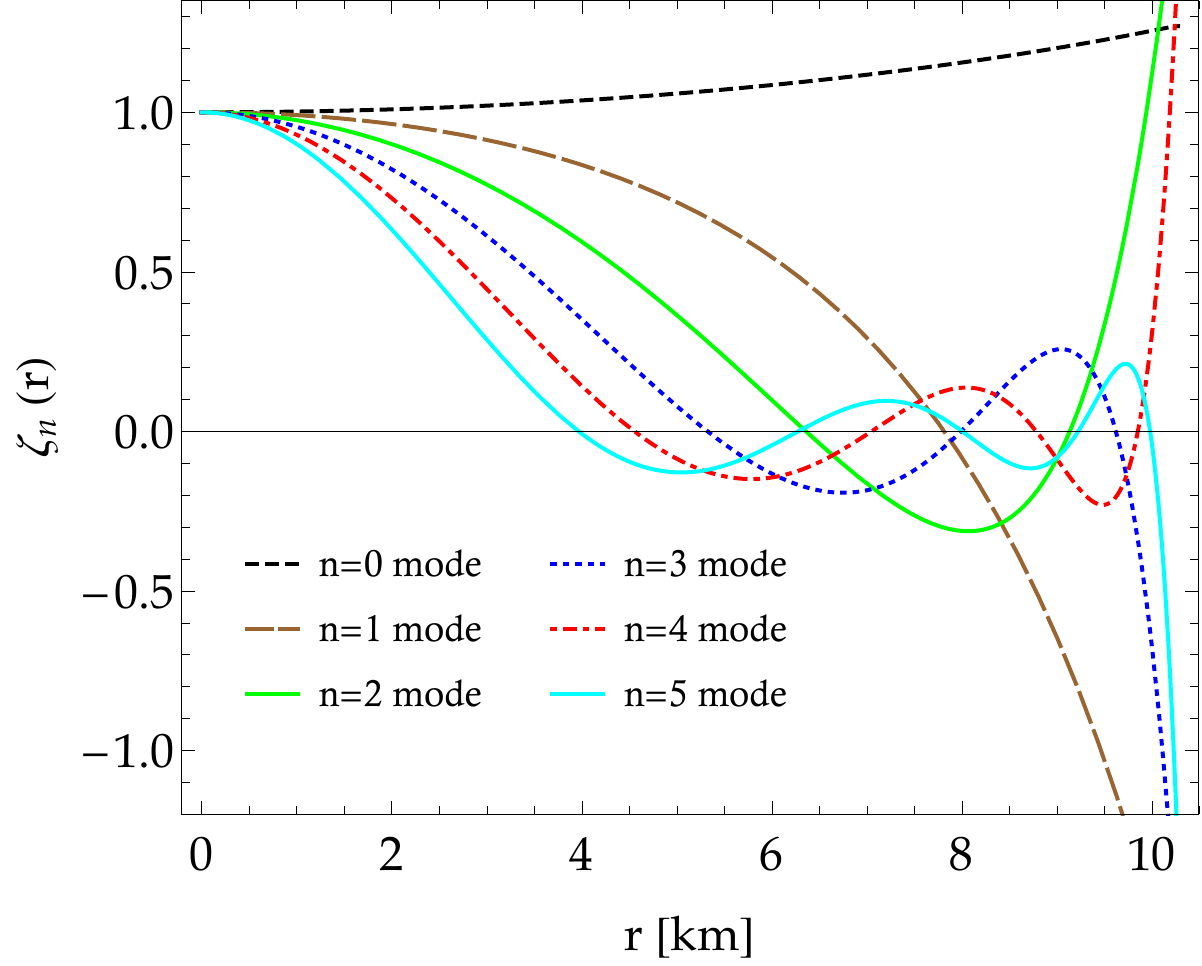}\
 \includegraphics[width=7.6cm]{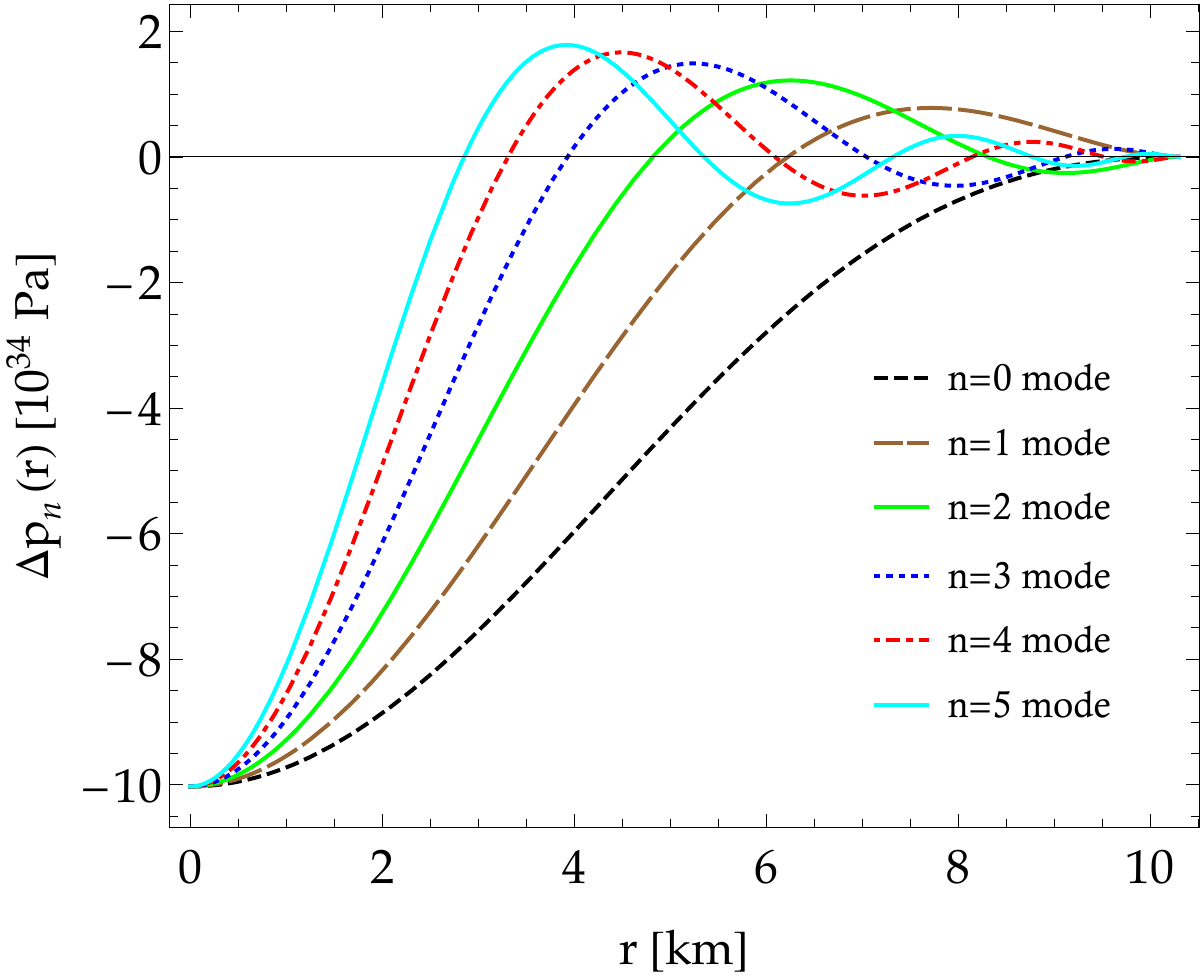}
 \caption{\label{figure6} Numerical solution of the radial pulsation equations (\ref{58}) and (\ref{59}) in the case of a neutron star with central density $\rho_c = 1.5 \times 10^{18}\ \rm{kg}/\rm{m}^3$ for $\alpha = 2.0 \times 10^{-6} \rm m^{-1}$. Left Panel: Eigenfunctions $\zeta_n(r)$ for the lowest six normal vibration modes which have been normalized at the center of the star. Right panel: Lagrangian perturbations of the pressure $\Delta p_n(r)$ where the boundary condition (\ref{61}) has been satisfied at the stellar surface. The corresponding values to the squared frequency of the different oscillation modes $\omega_n^2$ for such configuration are given in Fig. \ref{figure7}. }  
\end{figure}

\begin{figure}
\centering
 \includegraphics[width=7.8cm]{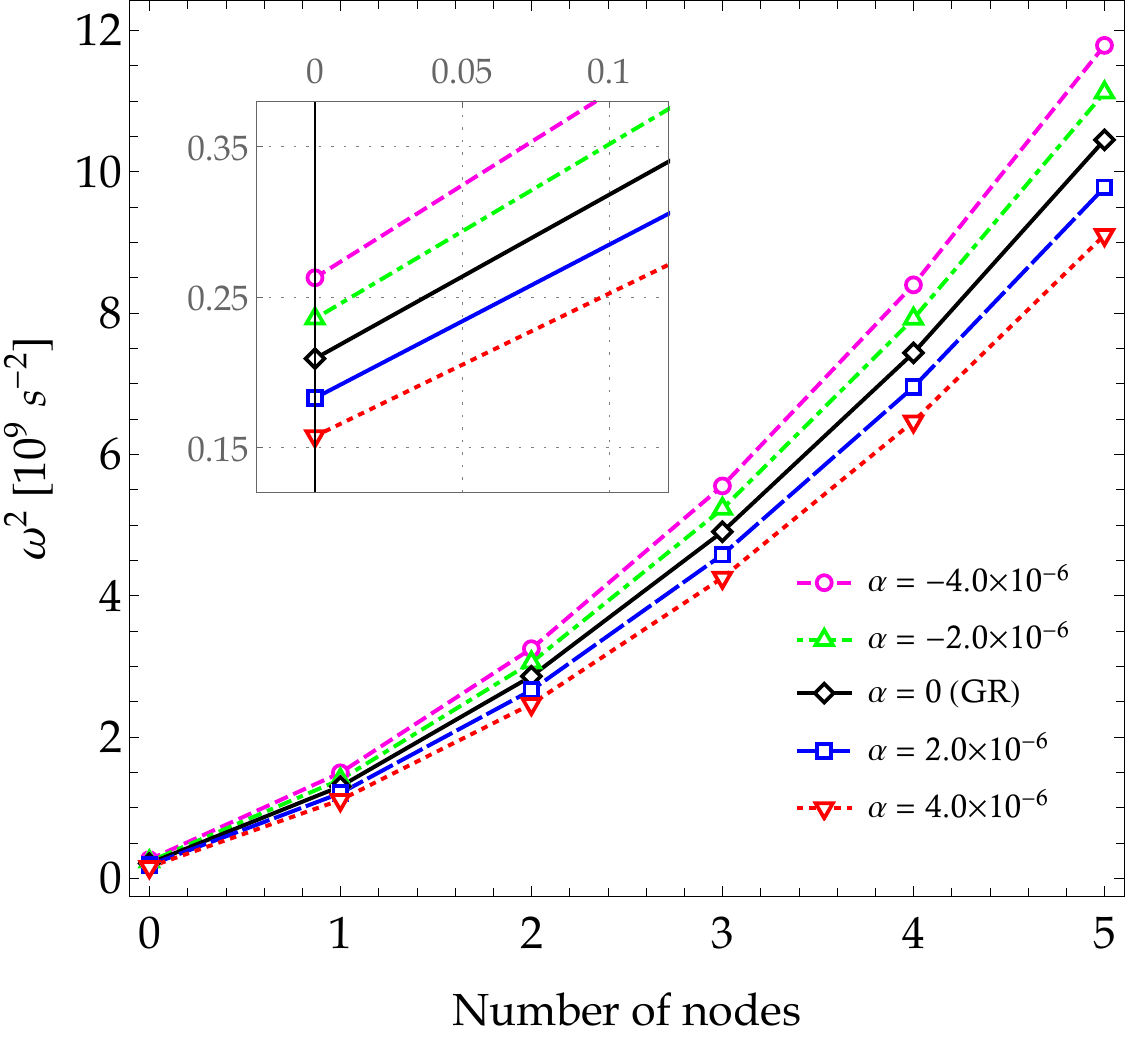}
 \caption{\label{figure7} Squared frequency of the pulsation modes against the number of nodes in the interior region of a neutron star with central density $\rho_c = 1.5 \times 10^{18}\ \rm{kg}/\rm{m}^3$, where five values of the parameter $\alpha$ are particularly considered. In the GR case we recover the frequencies $f_n = \omega_n/2\pi$ given in Ref. \cite{KokkotasR2001}.  }  
\end{figure}

Analogously as in GR, once the equilibrium quantities are known after solving the TOV equations, the numerical integration of oscillation equations (\ref{58}) and (\ref{59}) is carried out using the shooting method, that is, we integrate such equations for a set of trial values of $\omega^2$ satisfying the condition (\ref{60}). Moreover, we consider that normalized eigenfunctions correspond to $\zeta(0) = 1$ at the stellar center, and we integrate up to the  surface. Then the boundary condition (\ref{61}) will be fulfilled only for some values of $\omega^2$, which would correspond to the correct frequencies for each vibration mode. For instance, given a value of the parameter $\alpha$, the pulsation properties of the stellar model with central density $\rho_c = 1.5 \times 10^{18}\ \rm kg/m^3$ are shown in Fig. \ref{figure6}. Consequently, for a given configuration there are different eigenvalues $\omega_0^2 < \omega_1^2 < \cdots <\omega_n^2 < \cdots$ with their respective eigenfunctions $\zeta_n(r)$ and $\Delta p_n(r)$, where $n$ stands for the number of nodes inside the star. Thus, the fundamental mode is represented by $n=0$ (which has no nodes), whereas the first overtone $(n=1)$ has a node, the second overtone $(n=2)$ has two, and so forth. In Fig. \ref{figure7} we further investigate the dependence of the coupling parameter $\alpha$ on the vibration mode frequencies. Our results reveal that the greater the number of nodes between the center and the surface of the star, the oscillation frequency moves more rapidly away from its GR counterpart.

\begin{figure}
\centering
   \includegraphics[width=7.6cm]{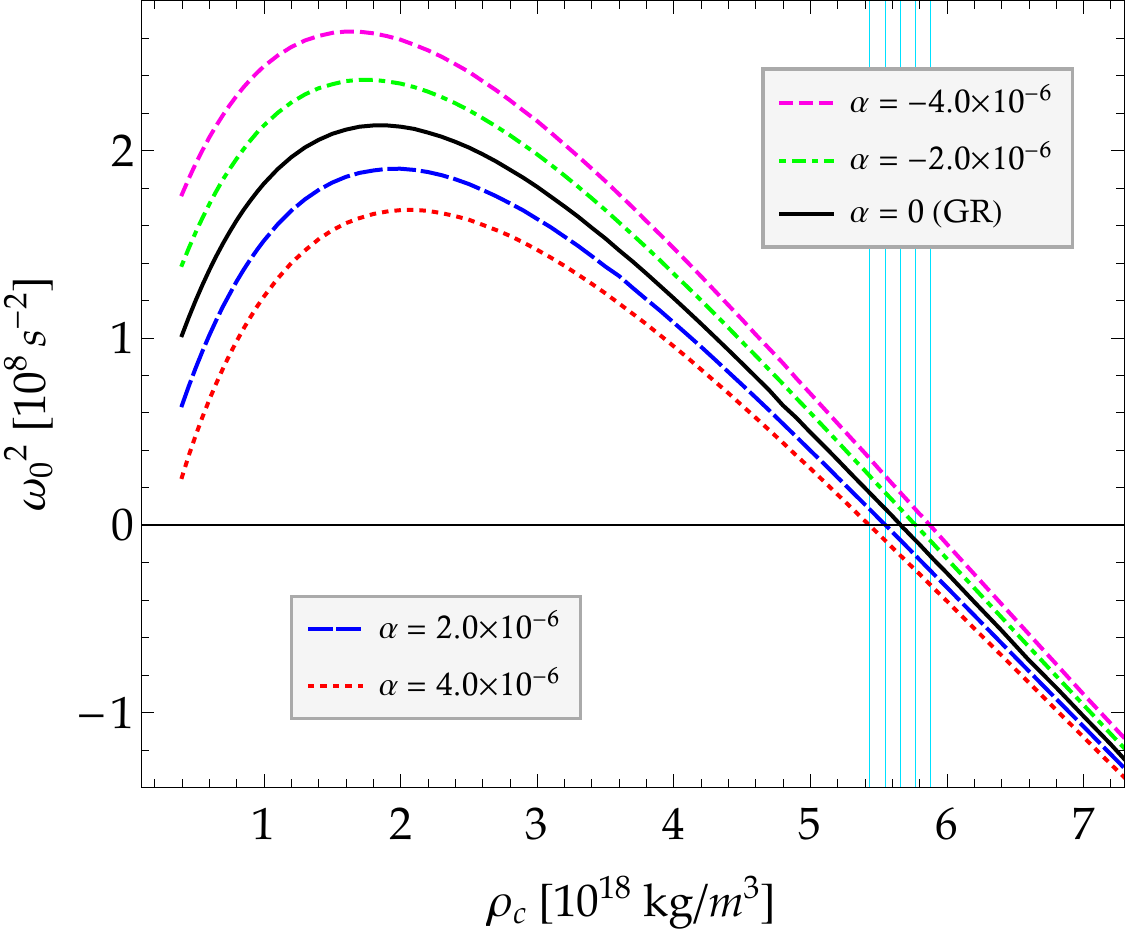}\
   \includegraphics[width=7.6cm]{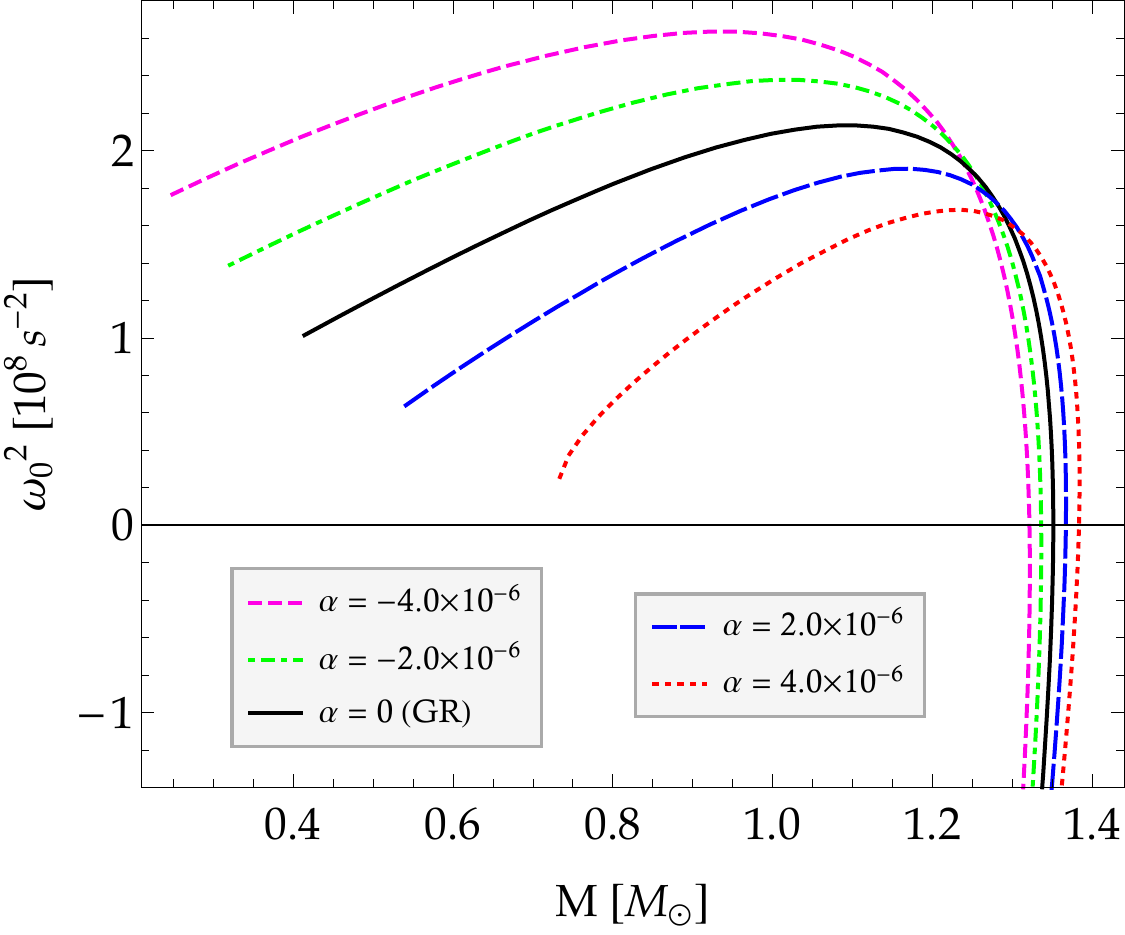}
   \caption{\label{figure8} Left panel: Squared frequency of the fundamental vibration mode versus central mass density in $f(R,T)= R+ h(T)$ gravity for the EoS I with different values of the parameter $\alpha$. The vertical lines in cyan indicate the central density values where $\omega_0^2= 0$, whose values precisely correspond to the minimum-binding-energy points on the right plot of fig. \ref{figure4}. Right panel: Squared frequency of the fundamental oscillation mode as a function of the total gravitational mass. }  
\end{figure}

Given the EoS I and a value for the parameter $\alpha$, we can now study the radial stability of the equilibrium configurations presented in Fig. \ref{figure2} by calculating the squared frequency of the fundamental mode. Figure \ref{figure8} gives a plot for the pulsation frequency as a function of central density (left panel) and a plot for the squared frequency of the fundamental mode against the total gravitational mass (right plot). From the left plot we can observe that $\omega_0^2(\rho_c)$ passes through zero at the central mass density value corresponding to the minimum binding energy configuration as shown in the right plot of Fig. \ref{figure4}. This means that the cusp formed by the binding energy as a function of the proper mass can be used to indicate the onset of instability in the case of relativistic polytropic stars in $f(R,T)= R+ h(T)$ gravity with conserved energy-momentum tensor. 

The critical central density corresponding to the maximum mass configuration deviates slightly from the central density value corresponding to the minimum binding energy. Nevertheless, the concept of gravitational binding energy has been shown to be useful in constructing stable equilibrium configurations since it is compatible with the calculation of oscillation mode frequencies.

\section{Concluding remarks} \label{Sect6}

In this paper we have investigated the physical features of compact stars within the context of $f(R,T)= R+ h(T)$ gravity with conserved energy-momentum tensor. By adopting a polytropic EoS the function $h(T)$ can assume a specific form so that $\nabla_\mu T^{\mu\nu} =0$, where a parameter $\alpha$ arises and its role is to control the deviations with respect to Einstein's gravity. We studied the effects of the $h(T)$ term on the neutron star structure after having numerically integrated the modified TOV equations. Such a term generates an extra mass contribution which has been interpreted as an effective mass function. As a result, the total gravitational mass of a star increases (decreases) for positive (negative) values of $\alpha$. We must also point out that the modifications of the mass-radius relation are more pronounced in the low-central-density region. This qualitative behavior is similar to the case of compact stars with non-conserved energy-momentum tensor as shown in Refs. \cite{Lobato2020, Pretel2021}. 

The behavior of the gravitational binding energy as a function of the proper mass reveals the formation of a cusp for the different values of $\alpha$. In addition, following the Chandrasekhar perturbative approach, we have addressed the dynamical stability problem through adiabatic radial oscillations and we derived the pulsation equations. By numerical solution of this eigenvalue problem, the vibration frequencies have been calculated for a wide variety of stellar configurations. Indeed, our results show that the minimum-binding-energy points can be used to determine the onset of radial instability because such points correspond to a critical central density where the squared frequency is zero. The analysis of the normal vibration modes leads to another important characteristic: the onset of instability is reached at a lower and lower central density value as the parameter $\alpha$ increases. Moreover, the grater the number of nodes between the center and the surface of a star, the pulsation frequency moves further and further away from its GR value.

The perturbative approach developed here for the adiabatic radial pulsations can be applied to other more realistic equations of state for the neutron-star matter. In fact, it is possible to use piecewise polytropes to describe more realistic neutron stars \cite{Read2009}. It would also be interesting to explore our analysis in the presence of an interface discontinuity where it has been shown (at least in GR) that there exist stable configurations with a central mass density exceeding the critical density corresponding to the largest gravitational mass \cite{Clemente2020, Pereira2018}. Definitively, this subject would require new junction conditions that hold at the interface \cite{Rosa2021} and should be left for future works.

\acknowledgments

JMZP acknowledges Brazilian funding agency CAPES for PhD scholarship 331080/2019. JDVA thanks to Universidad Privada del Norte and Universidad Nacional Mayor de San Marcos for the financial support - RR Nº$\,005753$-$2021$-R$/$UNMSM under the project number B$21131781$.


\end{document}